\documentclass[journal]{IEEEtran}

\ifCLASSINFOpdf
\else
   \usepackage[dvips]{graphicx}
\fi
\usepackage{url}

\usepackage{booktabs}
\usepackage{multirow}
\usepackage{graphicx}
\usepackage{orcidlink}
\usepackage{cite}
\usepackage{amsmath}
\everymath{\displaystyle}
\usepackage{xcolor}

\begin{document}

\title{RetinexGuI: Retinex-Guided Iterative Illumination Estimation Method for Low Light Images}

\author{Yasin Demir\textsuperscript{\orcidlink{0000-0002-0834-2780}}, Nur Hüseyin Kaplan\textsuperscript{\orcidlink{0000-0002-4740-3259}}, Sefa Kucuk\textsuperscript{\orcidlink{0000-0002-0279-3185}}, Nagihan Severoglu\textsuperscript{\orcidlink{0000-0002-3524-2566}}

\thanks{Yasin Demir, Nur Hüseyin Kaplan, Sefa Kucuk, and Nagihan Severoglu are with the Department of Electrical and Electronics Engineering, Erzurum Technical University, Erzurum, Turkey, 25050 (e-mails: yasin.demir@erzurum.edu.tr; huseyin.kaplan@erzurum.edu.tr; sefa.kucuk@erzurum.edu.tr; nagihan.severoglu@erzurum.edu.tr).}}

\markboth{}
{Shell \MakeLowercase{\textit{et al.}}: Bare Demo of IEEEtran.cls for IEEE Journals}
\maketitle

\begin{abstract}
In recent years, there has been a growing interest in low-light image enhancement (LLIE) due to its importance for critical downstream tasks. Current Retinex-based methods and learning-based approaches have shown significant LLIE performance. However, computational complexity and dependencies on large training datasets often limit their applicability in real-time applications. We introduce RetinexGuI, a novel and effective Retinex-guided LLIE framework to overcome these limitations. The proposed method first separates the input image into illumination and reflection layers, and iteratively refines the illumination while keeping the reflectance component unchanged. With its simplified formulation and computational complexity of $\mathcal{O}(N)$, our RetinexGuI demonstrates impressive enhancement performance across three public datasets, indicating strong potential for large-scale applications. Furthermore, it opens promising directions for theoretical analysis and integration with deep learning approaches. The source code will be made publicly available at https://github.com/etuspars/RetinexGuI once the paper is accepted.


\end{abstract}

\begin{IEEEkeywords}
Iterative enhancement, low-light image enhancement, real-time applications, Retinex theory. 
\end{IEEEkeywords}

\IEEEpeerreviewmaketitle

\section{Introduction}

\IEEEPARstart{A}{s}   real-time imaging systems become more common, capturing reliable images in low-light conditions is increasingly important. However, these images often suffer from reduced visibility and amplified noise, which can degrade downstream vision tasks. Therefore, the development and implementation of fast and efficient low-light image enhancement (LLIE) methods are crucial for improving visual quality and supporting practical real-time applications.


Currently, LLIE research is largely dominated by two main approaches: Retinex-based methods \cite{srie, 7782813, 8304597, 9032356, 9831058, 9738463, 9844872, 10106032, hu2025low, CHEN2026110316} and learning-based methods \cite{9328179, Liu_2021_CVPR, 9334429, 9369102, Ma_2022_CVPR, Cai_2023_ICCV, 10174279, 10154056, 10557144, 10623210,Morawski_2024_CVPR,  10663248, WEN2025111033, 10974676, HU2026110125}. Retinex-based methods model an observed image as the product of illumination and reflectance components under the Retinex assumption \cite{land1977retinex}. Early versions, including single-scale Retinex \cite{557356} and multi-scale Retinex \cite{597272}, treat the estimated reflectance as the final enhanced output. However, the results produced by these methods are often prone to over-enhancement and color distortion, leading to visually unnatural appearances. Recent Retinex-based approaches formulate LLIE as a variational optimization problem with explicit regularization on both components \cite{9032356, 10106032, hu2025low, CHEN2026110316}. These variational Retinex models typically rely on hand-crafted regularization terms and require careful parameter tuning, which increases computational cost and limits their real-time applicability. Learning-based algorithms, on the other hand, leverage deep neural networks to learn complex enhancement mappings from data, covering both learning-driven \cite{Ma_2022_CVPR, 9334429, 9369102, 10154056} and Retinex-inspired approaches \cite{Liu_2021_CVPR, Cai_2023_ICCV, 10974676}. Although they can produce impressive visual results, such models typically require large training datasets, substantial computational resources, and task-specific training, which limit their flexibility and suitability for real-time applications.

To overcome the aforementioned limitations, we introduce a novel Retinex-guided LLIE framework that provides fast and stable enhancement with minimal computational cost. In contrast to prior Retinex-based methods, we present a lightweight and computationally efficient enhancement strategy that iteratively refines the illumination component while keeping the reflectance unchanged, without relying on any external prior or complex regularization.

We can list the key contributions of the RetinexGuI as follows:

\begin{itemize}
    \item \textit{A simplified formulation:} Compared to recent Retinex-based approaches, our framework employs a simplified Retinex-guided formulation and does not require external priors or hyperparameter tuning.     
    \item \textit{Computational efficiency:} The proposed method eliminates the need for extensive training datasets or heavy computational resources. This provides a good balance between efficiency and visual quality, supporting real-time performance for various low-light enhancement tasks.
    \item \textit{Robust and versatile enhancement behavior:} Our RetinexGuI performs reliably in various low-light scenarios and offers potential for extension to related applications, such as virtual exposure generation.
\end{itemize}

\section{Proposed RetinexGuI Method}
\label{sec:proposed}

Based on Retinex theory \cite{land1977retinex}, an observed image $\mathbf{I}$ is expressed in terms of its illumination and reflectance components as

\begin{equation} 
\mathbf{I} = \mathbf{L} \odot \mathbf{R}
\label{equation1}
\end{equation}

\noindent here, $\mathbf{L}$ and $\mathbf{R}$ represent the illumination and reflectance components, respectively, and $\odot$ indicates element-wise multiplication. For convenience, we reformulate the Retinex model in the logarithmic domain as

\begin{equation} \label{equation2}
\displaystyle
   \mathbf{I}_l = \mathbf{L}_l + \mathbf{R}_l
\end{equation}

\noindent where $\mathbf{I}_l = \log(\mathbf{I})$, $\mathbf{L}_l = \log(\mathbf{L})$, and $\mathbf{R}_l = \log(\mathbf{R})$. We iteratively enhance the input image by updating the illumination component while keeping the reflectance component unchanged. Specifically, the enhanced image in the logarithmic domain is obtained as

\begin{equation} \label{equation3}
\mathbf{I}_l^{(N+1)} = \mathbf{L}_l^{(N)} + \mathbf{R}_l,
\end{equation}

\noindent where $N$ denotes the iteration number. The initial values are set as $\mathbf{I}^{(0)} = \mathbf{I}$ and $\mathbf{R}_l = \mathbf{I}^{(0)}$. 
The illumination component is then iteratively estimated by dividing the observed image by the initial reflectance component as



 
\begin{equation} \label{equation4}
    \mathbf{L}_l^{(N)} = \frac{\mathbf{I}_l^{(N)}}{\mathbf{R}_l}. 
\end{equation}

\noindent Given that $\mathbf{R}_l = \mathbf{I}^{(0)}$, we can rewrite (\ref{equation4}) as

\begin{equation} \label{equation5}
    \mathbf{L}_l^{(N)} = \dfrac{\mathbf{I}_l^{(N)}}{\mathbf{I}_l^{(0)}}. 
\end{equation}

\noindent By substituting (\ref{equation5}) into (\ref{equation3}), the update equation can be equivalently expressed as

\begin{equation} \label{equation6}
    \mathbf{I}_l^{(N+1)} = \dfrac{\mathbf{I}_l^{(N)} }{\mathbf{I}^{(0)}_l} + \mathbf{I}^{(0)}_l. 
\end{equation}

\noindent The recursive update given in (\ref{equation6}) is equal to the following: 

\begin{equation} \label{equation7}
\mathbf{I}_l^{(N+1)} 
= \frac{\sum_{n=0}^{N} \left(\mathbf{I}_l^{(0)}\right)^n}
       {\left(\mathbf{I}_l^{(0)}\right)^{(N)}}
  + \mathbf{I}_l^{(0)}
= \sum_{n=0}^{N} \left(\mathbf{I}_l^{(0)}\right)^{-n}
  + \mathbf{I}_l^{(0)} .
\end{equation}


\noindent When the number of iterations becomes sufficiently large  ($N \to \infty$), the formulation converges as follows:

\begin{equation}\label{equation8}
\mathbf{I}_l^{(\infty)} 
= \frac{1}{1-\left(\mathbf{I}_l^{(0)}\right)^{-1}} + \mathbf{I}_l^{(0)}
= \frac{\left(\mathbf{I}_l^{(0)}\right)^2}{\mathbf{I}_l^{(0)}-1}.
\end{equation}

\noindent It should be noted that all operations are performed on an element-wise basis. The closed-form result demonstrates that iterative illumination refinement converges to a solution defined by the initial illumination estimate. As the number of iterations increases, the enhancement process becomes independent of intermediate updates and is managed by a simple analytical formulation. This property explains the stable behavior of the proposed RetinexGuI and its ability to achieve consistent enhancement results with only a few iterations.


The result of (\ref{equation8}) is considered as the initial enhancement level. To obtain further levels of enhancement, especially for extremely dark scenes, (\ref{equation8}) is applied in a cascaded manner.


\section{Experiments and Results}

We perform all experiments on a PC equipped with Windows 11 OS, Intel Core i7-11800H CPU at 2.3 GHz, 16 GB of RAM, Nvidia RTX 3060 6 GB GPU, and MATLAB2024b\color{black}. We compare the RetinexGuI with five related methods, including SRIE \cite{srie}, Hu $et~al.$ \cite{hu2025low}, Wang $et~al.$ \cite{wang2023}, GCP-MCS \cite{gcpmcs}, and ACGC \cite{acgc}. For a fair comparison, we obtain the implementation codes from the authors’ official websites and follow the recommended parameter settings. We evaluate the performance of competing methods on three well-known low-light datasets: VE-LOL-L \cite{velol}, RELLISUR \cite{rellisur}, and LoLi-Street \cite{loli}. Among them, the VE-LOL-L dataset consists of 1000 synthetic paired and 500 real paired images; the RELLISUR dataset includes 85 paired images, each with five different exposure levels; the LoLi-Street dataset contains 3k paired images. 

\subsection{Implementation Details}

We first transform the input RGB-color low-light image into the Hue-Saturation-Value (HSV) space. Since the V channel represents illumination and reflectance information, we perform iterative illumination estimation on the V component. To handle under-exposed or extra-dark images, we repeatedly apply the converged illumination expression in (\ref{equation8}). 
We categorize the input images into three groups based on the mean value $(\mu_V)$ of the V component. 
The number of levels $(K)$ is determined as

\begin{equation}
\begin{aligned}
K =
\begin{cases}
1, &  \mu_V > 0.16, \\
2, &  0.08 \leq \mu_V \leq 0.16, \\
3, &  \mu_V < 0.08.
\end{cases}
\end{aligned}
\end{equation}

\noindent The converged illumination update in (\ref{equation8}) is then applied iteratively as


\begin{equation}
\begin{aligned}
\mathbf{V}_l^{k+1} &= \frac{\left(\mathbf{V}_l^{k}\right)^2}{\mathbf{V}_l^{k} - 1},
\qquad k = 1, 2, \ldots, K.
\end{aligned}
\end{equation}

\noindent Moreover, to provide more natural and balanced color reproduction, we apply a gamma correction to the S component using an empirical gamma value of 0.7 as

\begin{equation}
\mathbf{S}' = \mathbf{S} \odot \mathbf{S} ^{\gamma},
\end{equation}

\noindent and the enhanced HSV image is finally converted back to the RGB color space.


\subsection{Performance Comparison}


\subsubsection{Qualitative Comparison}

\begin{figure*} [htb!]
	\centering
	\includegraphics[width=1\textwidth]{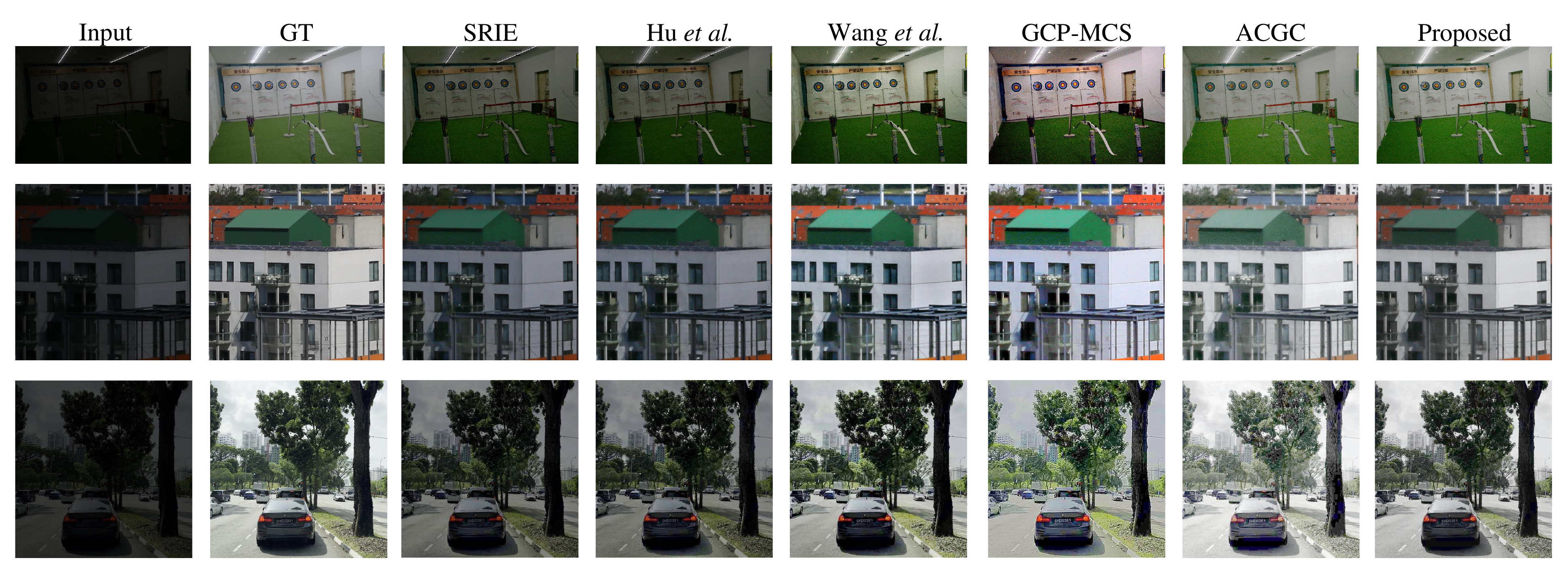}
	\caption{Visual comparison of low-light image enhancement results produced by different methods on three representative images from the three different datasets.}
	\label{fig:visual_results}
\end{figure*}

We perform visual comparisons between various enhancement techniques and the proposed RetinexGuI, as shown in Fig.~\ref{fig:visual_results}. 
As can be seen in Fig.~\ref{fig:visual_results}, SRIE and  Hu $et~al.$ produce insufficient enhancement, whereas GCP-MCS and ACGC exhibit noticeable color distortions and unnatural visual artifacts. Wang $et~al.$ fails to provide sufficient enhancement in local regions. Compared to other methods, the proposed RetinexGuI shows visually clear content with more natural enhancement results.


\subsubsection{Quantitative Comparison}

To demonstrate the effectiveness of our RetinexGuI, we use the peak signal-to-noise ratio (PSNR) and the structural similarity index measure (SSIM) \cite{ssim}. The higher values of PSNR and SSIM indicate better performance. Table \ref{Table1} reports the average PSNR and SSIM values of six competitive methods on three datasets. It can be seen from Table \ref{Table1} that the proposed RetinexGuI has the highest evaluation
scores among six methods.

To further evaluate the computational efficiency of the proposed RetinexGuI, we provide the processing time of the competing methods in Table \ref{Table:runtime}. We measure the average execution time (AET) on a set of 20 images sized $625\times625\times3$ from the RELLISUR dataset. As shown in Table~\ref{Table:runtime}, our method achieves the best runtime performance, as its computational complexity $\mathcal{O}(N)$ allows efficient implementation. Here, $N$ stands for the total number of pixels in the input image.

\begin{table} [t!]
\centering
\caption{Comparison of average PSNR and SSIM on three datasets. Bold and underlined scores denote the best and second-best results, respectively.}
\label{Table1}

\resizebox{\columnwidth}{!}{%
\begin{tabular}{llccc}
\toprule
 &  & \multicolumn{3}{c}{Datasets} \\
\cmidrule(lr){3-5}
Methods & Metrics & VE-LOL-L & RELLISUR & LoLi-Street \\
\midrule

\multirow{2}{*}{SRIE \cite{srie}} & PSNR & 13.601 & 12.344 & 16.213 \\
                      & SSIM & 0.403  & 0.431  & 0.819  \\

                      \midrule

\multirow{2}{*}{Hu et al. \cite{hu2025low}} & PSNR & 13.799 & 12.808 & 16.607 \\
                           & SSIM & \underline{0.408} & 0.451  & \underline{0.831}  \\
                    
\midrule

\multirow{2}{*}{Wang et al. \cite{wang2023}} & PSNR & 13.012 & 15.527 & 17.483 \\
                            & SSIM & 0.384  & 0.486  & 0.821  \\
                                      
\midrule

\multirow{2}{*}{GCP-MCS \cite{gcpmcs}} & PSNR & \underline{14.453} & 16.181 & \underline{18.937} \\
                         & SSIM & 0.381  & 0.469  & 0.810  \\
                                   
\midrule

\multirow{2}{*}{ACGC \cite{acgc}} & PSNR & 13.959 & \underline{17.404} & 18.486 \\
                      & SSIM & 0.395  & \underline{0.528}  & 0.808  \\

\midrule

\multirow{2}{*}{Proposed} & PSNR & \textbf{15.274} & \textbf{18.033} &\textbf{24.177}  \\
                          & SSIM & \textbf{0.425}  & \textbf{0.564}  & \textbf{0.840}  \\
                                    
\bottomrule
\end{tabular}%
}
\end{table}
\begin{table}[t!]
	
	\centering
	\caption{Comparison of Average Execution Time (in seconds) for Competing Methods.}
	\resizebox{.95\linewidth}{!}{
			\begin{tabular}{cccccc}
			\toprule
			SRIE \cite{srie}&Hu et al. \cite{hu2025low}&Wang et a. \cite{wang2023}&GCP-MCS \cite{gcpmcs}&ACGC \cite{acgc}&Proposed \\ \cmidrule{1-6}
			
			5.809&2.552&0.765&\underline{0.091}&0.217&\textbf{0.036}
			
			\\ \bottomrule
	\end{tabular}}
	\label{Table:runtime}
 
\end{table}

\subsection{Ablation Study}

\begin{figure} [t]
\centerline{\includegraphics[width=\columnwidth]{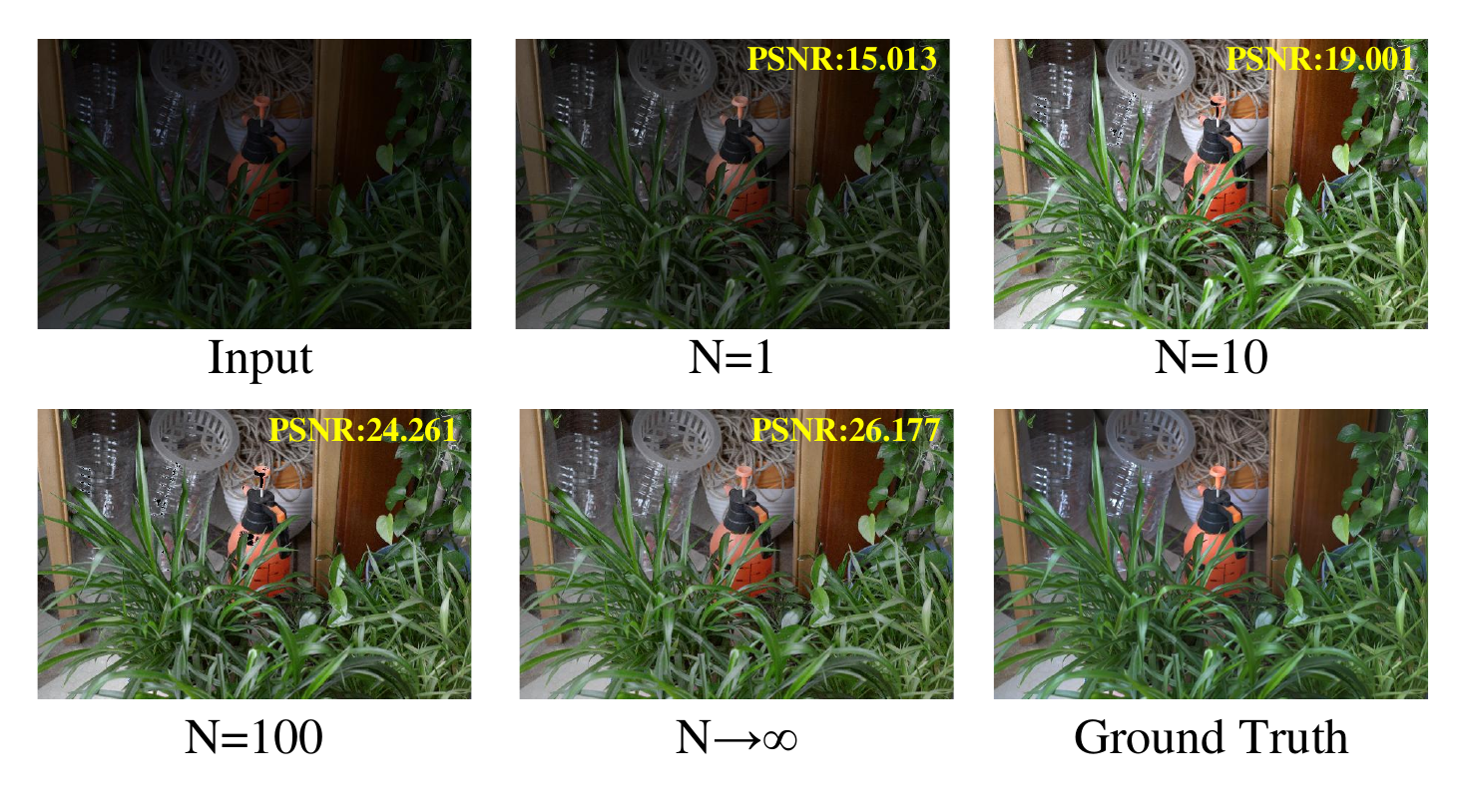}}
\caption{Visual comparison of using different numbers of iterations.}
\label{fig:figure1}
\end{figure}

To illustrate the effect of the number of iterations on the enhanced results of the proposed method, we conduct experimental studies by varying the number of iterations. From Fig.~\ref{fig:figure1}, we can note that increasing the number of iterations improves both visual quality and PSNR values. When we apply more iterations, the illumination becomes clearer while preserving color fidelity. Our RetinexGuI effectively enhances visibility in low-light images without noticeable color distortions. Additionally, with a sufficiently large number of iterations, the enhancement results stabilize and demonstrate the convergence behavior of the proposed RetinexGuI.


\section{Conclusion}

In this letter, we present a simple yet effective Retinex-guided framework for low-light image enhancement. The proposed RetinexGuI improves low-light images without relying on external prior information or complex parameter tuning. The experimental results demonstrate that the proposed method achieves competitive performance in terms of visual quality and quantitative evaluation metrics. Thanks to its simplified formulation, the proposed approach has negligible computational cost and is highly suitable for real-time low-light image enhancement.

This study demonstrates a significant advancement in low-light image enhancement by combining computational efficiency with strong enhancement performance. It provides a promising foundation for further theoretical investigation and integration with learning-based models. Future research should address robustness under extreme illumination conditions and extend the method to more challenging real-world low-light imaging scenarios.



\bibliographystyle{IEEEtran}
\bibliography{refs2025}

\end{document}